# Development of High Performance Electron Beam Switching System for Swiss Free Electron Laser at PSI

**Martin Paraliev, Christopher Gough**

Paul Scherrer Institute, 5232 Villigen PSI, Switzerland

## ABSTRACT

A compact X-ray Free Electron Laser (SwissFEL) is under development at the Paul Scherrer Institute. To increase facility efficiency the main linac will operate in two electron bunch mode. The two bunches are separated in time by 28 ns and sent to two undulator lines. The combination of two beam lines should produce short X-ray pulses covering wavelength range from 1 to 70 Å with submicron position stability. To separate the two bunches, a novel electron beam switching system is being developed. The total deflection is achieved with a combination of high Q-factor resonant deflector magnet, followed by a DC septum magnet. The shot-to-shot deflection stability of the entire switching system should be <+/-10 ppm in amplitude and +/-100 ps in time, values which present severe measurement difficulties. Deflection magnets requirements, development and results of the kicker prototype are presented.

## 1 INTRODUCTION

The Swiss X-ray Free Electron Laser [1] (SwissFEL) at the Power Scherrer Institute is a linear accelerator (linac) based 4$^{th}$ generation light source for applied science research. It will use high energy (up to 5.8 GeV) electron bunches to produce short (2 to 20 fs) and high brightness (up to $6.10^{35}$ photons.mm$^{-2}$.mrad$^{-2}$.s$^{-1}$) X-ray pulses covering the spectral range from 1 to 70 Å [2]. To increase facility efficiency the main linac should operate in two electron bunch mode. The two bunches are separated in time by 28 ns and at 3 GeV they are switched to two shorter linac and undulator lines for simultaneous operation. A novel approach using fast resonant kicker magnet is being explored in order to provide fast, reliable and high stability switching.

## 2 SYSTEM LAYOUT AND REQUIREMENTS

Design parameters of the switching system are summarized in Table 1. The switching system consists of a fast compound "kicker" magnet to provide fast small beam displacement followed by a septum magnet to provide large and stable beam deflection. In the present design three resonant kicker magnets followed by DC septum are used. A pulsed septum is considered as an option because of reduced dissipated power in the system.

**Table 1**. Switching system design parameters

| Parameter | Value |
|---|---|
| Electron beam energy | 3 GeV |
| Max. rise time | 28 ns |
| Beam size H (V) | 20 (50) µm rms |
| Deflection angle | 2° (35 mrad) |
| Deflection angle tolerance | ±10 ppm |
| Available length | 14 m |

From previous experience a beam separation at the septum entrance of 7 mm is a good compromise. Using the principle of equipartition of errors, the tolerances are divided equally between amplitude and time jitter, and inversely proportional to the deflection angle. The stability requirements for the kicker system and the septum are summarized in Table 2.

**Table 2**. Kicker and septum stability requirements

| Parameter | Kicker system | Septum |
|---|---|---|
| Considered technology | Resonant | DC |
| Pulse shape | Sine wave | DC |
| Total deflection | 1 mrad | 33 mrad |
| Jitter due to amplitude | ±2.5 ppm | ±5 ppm |
| Jitter due to timing | ±2.5 ppm | - |
| System amplitude stability | ±82.5 ppm | ±5 ppm |
| Time jitter | ±160 ps | - |

In order to fulfill these tight stability requirements a resonant kicker with solid state driver is considered. Initial stability measurement results from a prototype magnet are presented and discussed.

## 3 RESONANT KICKER MAGNET

The first of two basic kicker (short rise time) magnet types is the transmission line one. If strip line topology is used for a given clear beam aperture, an optimum transmission line impedance can be found to minimize power dissipation in the termination. Fig. 1. shows magnetic field on beam axis as a function of transmission line characteristic impedance for a clear beam aperture of 16 mm. For strip line impedance calculation, an approximation formula is used with error < 0.5% [3] and the surrounding space is left open. The magnetic field on axis is numerically simulated, giving an optimum impedance value of 132 Ω. For 3 GeV electron beam energy, a practical magnet length (1 m) and 1 mrad deflection about 260 A and 35 kV are needed.

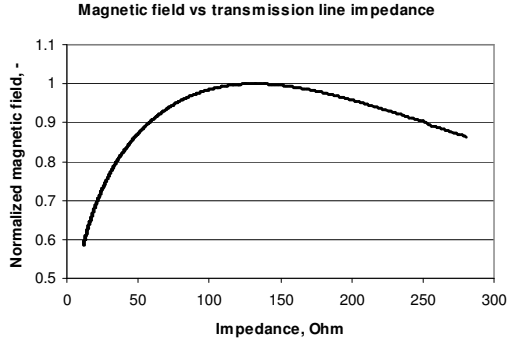

**Figure 1.** Normalized magnetic field on beam axis in function of transmission line impedance for fixed power dissipation in the termination.

The voltage would be difficult for semiconductor switch system and this would force the use of a less stable thyratron system.

Second kicker magnet type is a lumped inductor. In order to have a short rise time, a single turn inductor using parallel conductor bar (30 x 30 mm) with no magnetic cores is used. For the same deflection as the transmission line type, this will require current about 510 A. Inductance of such layout is in order of 300 nH. Adding the stray inductance of the switch and the connecting conductors the total inductance of the magnet could easily reach 600 nH. For the required rise time (28 ns) minimum voltage over perfect current source driver during transient is 11 kV.

The problem can be approached differently by considering the circuit time constants. For a RL circuit to settle to 80 ppm requires 9.4 time constants. For the lumped magnet a series resistance of 200 Ω is needed and the source voltage is in 100 kV class. If nonsteady state operation is chosen and practical time jitter $t_j$ of a semiconductor switch (1 ns) is taken for 80 ppm current stability ($E_a$), the acceptable current slope is:

$$\frac{dI}{dt} = \frac{I_{Nom} \cdot E_a}{t_j} = 4 \cdot 10^7 \text{ A/s} \quad (1)$$

Current in a transient RL circuit is described as:

$$I(t) = I_{Nom}[1 - \exp(-t/\tau)], \quad (2)$$

and its time derivative is:

$$\frac{dI}{dt} = \frac{I_{Nom}}{\tau} \exp(-t/\tau) \quad (3)$$

The tabulated values in Table 3 show the consequence of progressively larger time constants. With a time constant of 2.8ns, the steady state is reached rapidly and the switch time jitter plays no role in the final stability, but the needed source voltage is impractically high. However, when the circuit time constant is larger, the electron beam deflection occurs while the kicker magnet field is still rising. In this transient regime, the exact field is dependent upon the switch jitter, typically 1 ns.

**Table 3.** System parameters in steady state and transient regime

| RL Time Constant τ, ns | Total Series R, Ω | Field as proportion of steady state value | Required Source Voltage U, kV |
|---|---|---|---|
| 2.8 | 214.29 | 100% | 109.3 |
| 9.3 | 64.29 | 95% | 34.5 |
| 28.0 | 21.43 | 63.2% | 17.3 |
| 93.3 | 6.43 | 25.9% | 12.6 |
| 280.0 | 2.14 | 9.5% | 11.5 |
| 933.3 | 0.64 | 3.0% | 11.1 |
| 2800.0 | 0.21 | 1.0% | 11.0 |
| 9333.3 | 0.06 | 0.3% | 10.9 |
| 28000.0 | 0.02 | 0.1% | 10.9 |

It can be seen that the required voltage tends asymptotically to about 11 kV but the corresponding values of series resistance are impractically small.

Both magnet types described above can be used in an oscillatory mode. Usually, this option is not considered due to increased complexity. Nevertheless the oscillatory solutions has several advantages: the positive and negative half cycles can be both used to kick the electron bunches in the opposite direction reducing the needed current amplitude by factor of 2; the required energy can be put slowly in the system relaxing driver peak power requirements and working voltage; the system has a high Q factor, guaranteeing low phase noise and high stability; the option for narrow bandwidth, low noise measurements. These features of an oscillatory system and especially the lower driver voltage (solid state driver) were the driving force to explore a resonant kicker solution. Some of our previous work has shown that transmission line resonators at tens of MHz do not have such a high quality factor due to their large physical length and respectively large losses. In this article we have chosen to focus on a lumped, high Q, resonant, kicker magnet. The goal was to explore the possibilities to use compact solid state driven resonant magnet. Last but not least was the goal to explore the technology limitations in precision amplitude and phase measurements required for this scheme.

## 4 KICKER MAGNET PROTOTYPE

### 4.1 RESONATOR

Fig. 2 shows the constructed prototype resonant magnet made with copper parts. The first prototype used aluminium parts but with the development of precise amplitude measurements it was clear that the contact resistance between different parts is changing, causing irregular redistribution currents that compromised the stability measurements. The next prototype with copper parts increased the Q factor of the resonator and reduced the shot-to-shot amplitude jitter. For the capacitive element of the resonator, two 250 pF / 35 kV RF capacitors (CFMN-250EAC/35-DH-G) in parallel were used. The anticipated resonator voltages are around 10 kV peak. The resonator period of 56 ns was chosen to be twice the electron bunch separation. The Q factor of the system is 155.

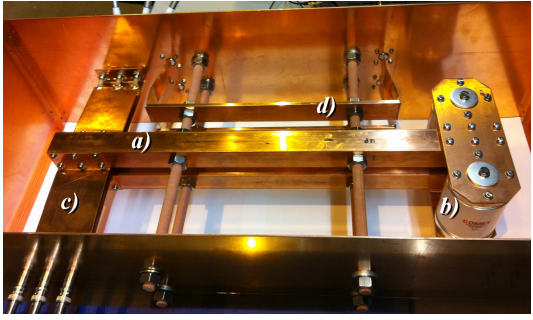

**Figure 2.** Resonator – *a)* conductor bars, *b)* RF capacitors, *c)* excitation strips and *d)* pickup

The driving scheme and the resonator layout are carefully chosen in order to keep the high voltage inside the magnet case and to minimize the unwanted parasitic resonant modes. Calibrated pickups are used to monitor the phase and the amplitude of the resonator. A set of six standard SHV connectors at the resonator lower side were chosen to form low inductance interface to the driver.

### 4.2 DRIVER

The driver consists of 6 driver modules in parallel. Each module has high voltage (1 kV), high speed MOS transistor (DE375-102N12A) and connects with individual coaxial cable. The transistors are heavily capacitively loaded to reduce the voltage swing over them to about 300 V with large enough margin to handle the burst turn-off transient. Each transistor has an individual high speed, high current (20 A) gate driver.

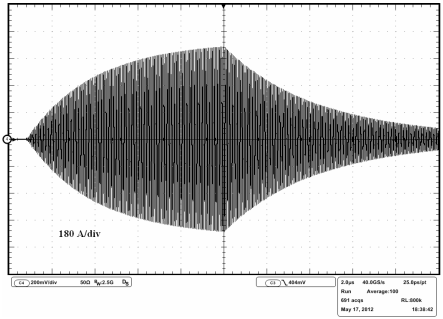

**Figure 3.** Typical resonant transient, (2us/div horizontal, 180A/div vertical)

Due to the difficult shot-to-shot stability requirements, special care was taken to ensure low phase noise through the whole driver chain. Despite the small average power both the transistors and the gate drivers are mounted on heat sinks to ensure small (controlled) temperature drift. Fig. 3 shows the resonator, driven for ~10 μs in order to reach the nominal current amplitude (500 A).

## 5 MEASUREMENT SCHEMES

### 5.1 PHASE

A precision phase measurement system was developed to measure the phase on cycle-by-cycle basis in preparation for rapid phase-locking to the rest of the accelerator. This was helpful to study the small time scale behavior of the system. The resolution limit of the system is better than 0.1° RF (15 ps) peak-to-peak. Due to the interaction of amplitude and phase, the precision amplitude measurement can also be used as an indicator for the phase stability of the system.

### 5.2 AMPLITUDE

To achieve cycle-by-cycle amplitude measurement at 10 ppm level, a "balanced measurement" method is used. It is based on measuring the difference between the measured sine wave and a DC offset. This scheme still requires a digitizing rate in the 12 Gs/s range but reduces the measurement dynamic range. A major drawback of the method is that the measurement system front-end should be able to withstand large overload for part of the cycle. Modern ADCs provide 8 bit resolution with LSB of 1 mV at conversion rates > 20 Gs/s but to achieve the required precision their front-ends should withstand 100 V without significant reduction of the bandwidth and fast (within few ns) recovery from deep saturation. Different large bandwidth limiter elements were evaluated in order to find a suitable solution. Table 4 compares the performance of several fast recovery limiting elements.

**Table 4.** Comparison of fast recovery limiting elements

| Limiting element | $t_{rr}$, ns | $C_o$, pF | $f_{cutoff}$, MHz | | $R_d$, Ω | $U_{limit}$, V |
|---|---|---|---|---|---|---|
| | | | Expected | Measured | | |
| BAS70-04 | 5 | 2 | 1592 | >1000 | 1.8 | 3.2 |
| BAT750 | 5 | 135 | 24 | 27 | 1.3 | 1.6 |
| BAT754S | - | 11 | 290 | 308 | 2.0 | 3.2 |
| BAV99 | 6 | 1.15 | 2769 | >1000 | 5.7 | 6.1 |
| BFG591[1] | NA | 5.4 | 590 | 530 | 4.3 | 4.6 |
| BFR93[1] | NA | 3.2 | 995 | 960 | 1.4 | 2.3 |
| SMP1307 | 1500[2] | 0.25 | 12739 | >1000 | <0[3] | 5.0 |
| SMP1330 | 4[2] | 0.7 | 4550 | >1000 | 2.1 | 2.8 |

[1] Transistors with base and collector shorted
[2] Carrier lifetime
[3] The element shows negative differential resistance

In Table 4, $t_{rr}$ is reverse recovery time, $C_o$ is device capacitance at zero bias, $f_{cutoff}$ is the cutoff frequency of the parasitically formed low pass filter on a 50Ω system, $R_d$ is the incremental resistance at heavy loading (0.4 - 1.0 A) and $U_{limit}$ is the limiting voltage over the device at 1 A.

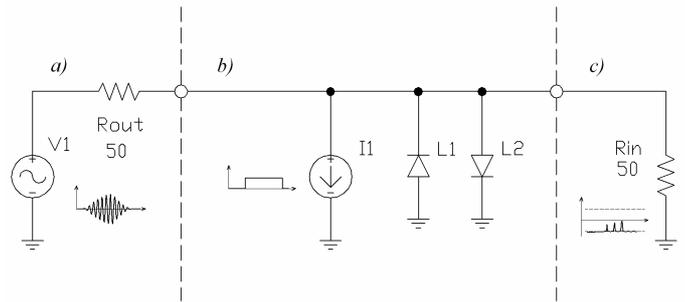

**Figure 4.** Simplified equivalent circuit of balanced measurement – *a)* voltage source with unknown amplitude, *b)* current balancing node with voltage limiter and *c)* measurement system.

An RF transistor, connected with its base and collector shorted, gave a reasonable compromise of low incremental resistance and wide bandwidth. With this pin connection, the transistor is kept in linear regime as much as possible. This allows the transistor to turn-off quickly after the overload. The transistor type BFR93 was chosen.

The other important element is generating the DC offset. To avoid working with high voltages, the DC offset is provided by a stable current source as shown in Fig. 4. Pulsed operation of the current source is needed to reduce the average power dissipation in the limiting element. Fig. 5 shows the typical waveform coming out of the balancing circuit.

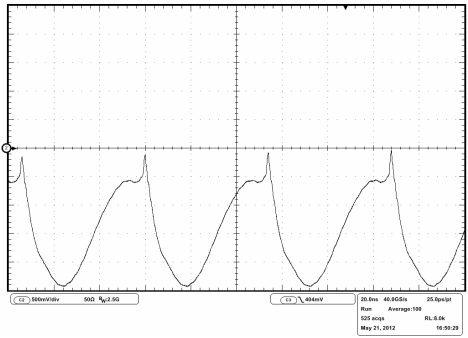

**Figure 5.** Signal waveform after "balancing", the small upper peaks give the wanted measurement near to zero volts, the lower "sine" shapes are from finite incremental resistance of the limiter.

Only the tip of the sine wave goes through the balancing circuit without distortion. The output voltage amplitude is limited to about ±2.5 V and the measured peak is kept around zero. Assuming a useful (linear) range of +/-200 mV the pulse duration (available sampling time) is only 560 ps. From Fig. 4, for 80 V sine wave, the expected sensitivity of the system is 25 ppm/mV.

## 6 MEASUREMENTS

With the balanced measurement system, the shot-to-shot stability of 100 sequential pulses was measured at 610 A. The result is shown in Fig. 6. The calibrated system sensitivity was found to be 32 ppm/mV. The difference of sensitivity from the expected value is attributed to the parasitic capacitance of the current source and limiting elements. The measured shot-to-shot stability is 53 ppm peak-to-peak. Subtracting the measurement noise floor linearly (22 ppm), the measured system stability is 31 ppm peak-to-peak.

Cycle-by-cycle amplitude change due to phase variation was also measured. One of the driving pulses was deliberately shifted with 1° RF cycle (150 ps). This single cycle phase change gave an amplitude change of the affected cycle of 470 ppm (Fig. 7). Thus amplitude change due to phase change is described with a sensitivity of 3.1 ppm/ps.

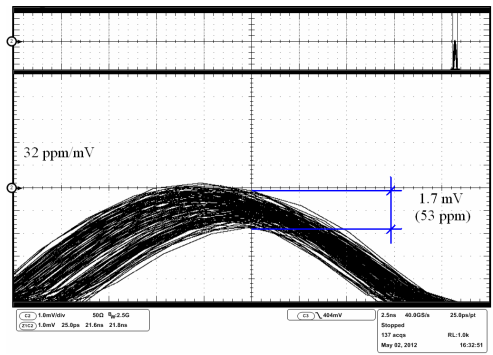

**Figure 6.** Signal waveform after balancing.

Assuming that the measured amplitude fluctuation of 31 ppm is entirely due to phase jitter, the expected phase noise of the driver should be less than the exceptionally small value of 0.07° (10 ps).

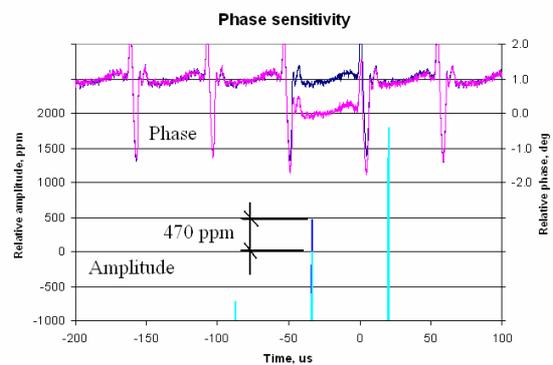

**Figure 7.** Cycle-by-cycle amplitude change due to deliberately introduced phase shift.

## 7 CONCLUSION

For double bunch operation of SwissFEL, fast (< 28 ns) high shot-to-shot stability (< 10 ppm peak-to-peak) beam switching system is being developed. A resonant kicker scheme combined with a DC septum magnet has been chosen. A prototype kicker magnet, driver and diagnostics system were designed and built. The preliminary measurements of the system show that it is feasible to reach the required stability. Measured shot-to-shot amplitude stability of the kicker is 31 ppm (peak-to-peak). For the considered layout this value translates to 1 ppm peak-to-peak stability of the whole switching system (kicker instability contribution). It was found that the resonant kicker amplitude is sensitive to phase noise. Special care was taken to reduce driver phase noise.

The next step is to prove the long term stability of the system including phase and amplitude feedback loops.

## REFERENCES

[1] http://www.psi.ch/swissfel/
[2] SwissFEL Conceptual Design Report, PSI Bericht Nr.10-04, April 2012
[3] Sophocles J. Orfanidis, "Electromagnetic Waves and Antennas", ch.10, p. 405, Rutgers University, NJ, 2008